\documentclass[showpacs,twocolumn,prl]{revtex4}
\usepackage{amsfonts}

\usepackage{amssymb}
\usepackage{graphicx}
\usepackage{dcolumn}
\usepackage{bm}
\begin{document}

\title{Theoretical prediction of topological insulator in ternary rare earth chalcogenides}
\author{Binghai Yan$^{1,2}$, Hai-Jun Zhang$^{2}$, Chao-Xing Liu$^{3}$,
Xiao-Liang Qi$^{4,2}$, Thomas Frauenheim$^1$ and Shou-Cheng Zhang$^2$ }

\affiliation{$^{1}$Bremen Center for Computational Materials Science,
Universit$\ddot{a}$t Bremen, Am Fallturm 1, 28359 Bremen, Germany\\
  $^{2}$Department of Physics, McCullough Building, Stanford University,
    Stanford, CA 94305-4045\\
$^{3}$Physikalisches Institut (EP3) and
  Institute for Theoretical Physics and Astrophysics,
  University of W$\ddot{u}$rzburg, 97074 W$\ddot{u}$rzburg, Germany\\
    $^{4}$Microsoft Research, Station Q, Elings Hall, University of
California, Santa Barbara, CA 93106, USA
}

\date{\today}

\begin{abstract}
A new class of three-dimensional topological insulator, ternary rare earth
chalcogenides, is theoretically investigated with {\it ab initio} calculations.
Based on both bulk band structure analysis and the direct calculation of
topological surface states, we demonstrate that LaBiTe$_3$ is a topological
insulator. La can be substituted by other rare earth elements, which provide
candidates for novel topological states such as quantum anomalous Hall
insulator, axionic insulator and topological Kondo insulator. Moreover,
YBiTe$_3$ and YSbTe$_3$ are found to be normal insulators. They can be used as
protecting barrier materials for both LaBiTe$_3$ and Bi$_2$Te$_3$ families of
topological insulators for their well matched lattice constants and chemical
composition.

\end{abstract}

\maketitle


{\it Introduction-} A topological insulator (TI) is a novel quantum state,
which has attracted great attention in condensed-matter physics recently
\cite{qi2010,moore2010,hasan2010}. TIs in two or three dimensions have both
insulating bulk energy gap and gapless edge or surface states on the sample's
boundary. The surface states of three-dimension (3D) TIs consist of odd number
of massless Dirac cones, which are protected by the time-reversal symmetry. TIs
can be most generally defined as a new state of quantum matter whose effective
electromagnetic action is given by the topological term
$S_{\theta}=(\theta/2\pi)(\alpha/2\pi)\int
\mathrm{d}^3x\mathrm{d}t\mathbf{E}\cdot\mathbf{B}$, with $\alpha$ the
fine-structure constant and the parameter $\theta=0$ or $\pi$ for trivial
insulator and TI, respectively \cite{qi2008}. Many novel properties have been
proposed as a result of this topological response \cite{qi2008,qi2009,li2010},
which are interesting in both fundamental physics and device applications. In
proximity with an ordinary superconductor, TI also provides a new candidate for
topological quantum computation \cite{fu2008}.

\begin{figure}
    \begin{center}
    \includegraphics[width=3.5 in]{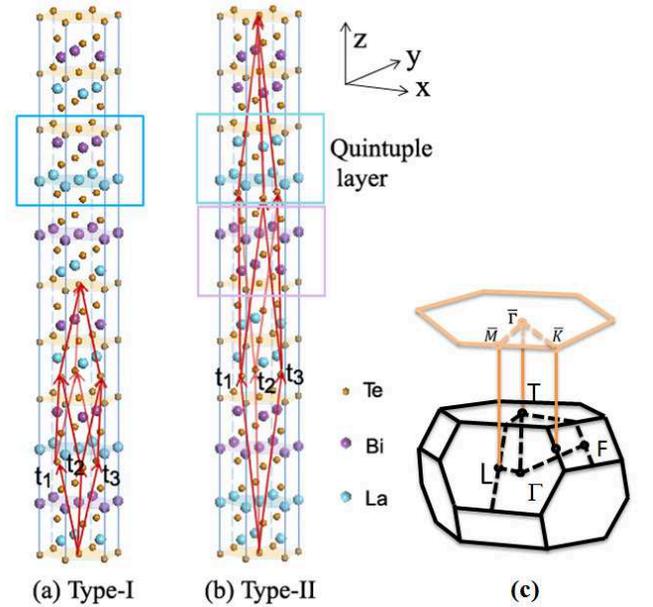}
    \end{center}
\caption{ (Color online) Crystal structures of LaBiTe$_3$ with three primitive
lattice vectors denoted as \textit{\textbf{t}}1;2;3 for (a) type-I and (b)
type-II lattices. The quintuple layers, Te-La-Te$^\prime$-Bi-Te for type-I
lattice, and Te-La-Te$^\prime$-La-Te and Te-Bi-Te$^\prime$$^\prime$-Bi-Te for
type-II lattice are indicated by rectangles. The $c$ axis of equivalent
hexagonal lattice is along the $z$ direction. (c) The first Brillouin zone is
shown for the rhombohedral lattice as well as the 2D projected surface. }
    \label{fig:crystal}
\end{figure}

The first TI was theoretically predicted and experimentally observed in the
HgTe quantum well\cite{bernevig2006d,koenig2007}. This work discovered the
basic mechanism of band inversion driven by spin-orbit coupling (SOC), which
serves as a template for most TIs discovered later. Since then, several TIs in
both 2D and 3D have been theoretically proposed and experimentally realized,
Bi$_{1-x}$Sb$_x$ alloy\cite{fu2007a,hsieh2008}, the family of Bi$_2$Se$_3$,
Bi$_2$Te$_3$ and Sb$_2$Te$_3$\cite{zhang2009,xia2009,chen2009}, the family of
TlBiTe$_2$ and TlBiSe$_2$\cite{yan2010,chen2010,sato2010direct}. Many other
materials are theoretically predicted and still waiting for experimental
verification\cite{chadov2010,lin2010a,lin2010b,xu2010,klintenberg2010,feng2010,jin2010}.
Of particular interest are the magnetic atom doped topological insulators for
the effects of magnetic impurities and ferromagnetism on the topological
surface states\cite{liu2009prl,Hor2010,yu2010}. In this letter, we report a new
family of ternary chalcogenides compound LnBT$_3$(Ln = rare earth elements, B =
Bi, Sb, and T = Te, Se), which share the same crystal structure with
Bi$_2$Se$_3$ family. LnBT$_3$ can be regarded as Bi$_2$Se$_3$ with half of the
Bi atoms replaced by Ln. Unlike Bi$_2$Se$_3$ family, in these materials
inversion symmetry is not always preserved. Since lanthanum(La) and Yttrium(Y)
have no $4f$ electrons, LaBiTe$_3$, YBiTe$_3$, LaSbTe$_3$ and YSbTe$_3$ are
four simplest ternary rare earth chalcogenides which have been synthesized in
experiments\cite{madelung2000}. In this work we focus on these four compounds
based on \textit{ab initio} calculations, among which we find that LaBiTe$_3$
is a strong topological insulator. Moreover, La and Y can be replaced by other
rare earth elements which have magnetic moments due to the localized $f$
electrons. These systems could provide realizations of the topological Mott
insulator state\cite{raghu2008}. The versatility of rare earth elements will
bring us new possibilities beyond Bi$_2$Te$_3$ family.

\begin{table*}
\caption{\label{table:structure} The experimental lattice
parameters\cite{madelung2000} $a$ and $c$ of hexagonal lattice in unit of
angstrom, the theoretically optimized atomic positions and band gaps. Type-II
structure has twice the length of type-I along $c$ axis (see Fig.1) and
thereafter has double atoms. The atomic coordinate is (0,0,$z_0c$) in the
rhombohedral primitive unitcell. We listed $z_0$ values for all atoms. These Te
atoms as inversion centers of type-II are marked as  Te$^\prime$ and
Te$^\prime$$^\prime$. The energy gap($E_g$) is in unit of eV.}
\begin{ruledtabular}
\begin{tabular}{c|rrrrr|rrrrr}
  &\multicolumn{4}{c}{Type-I} && \multicolumn{4}{c}{Type-II} &\\
  &LaBiTe$_3$&YBiTe$_3$&LaSbTe$_3$&YSbTe$_3$&&LaBiTe$_3$&YBiTe$_3$&LaSbTe$_3$&YSbTe$_3$&\\
  \hline
  $a$ &4.39&4.46 &4.24&4.47 && 4.39  &4.46  &4.24 &4.47&\\
  $c$ &30.20&31.65&30.40&30.32&&60.40  & 63.30&60.80&60.64&\\
  Te(Te$^\prime$/Te$^\prime$$^\prime$)& 0.0 & 0.0 & 0.0 & 0.0&&0.0/0.5&0.0/0.5&0.0/0.5&0.0/0.5&\\
  Te&0.20876&0.22243&0.20426&0.21850&&$\pm~$0.10604&$\pm~$0.10764&$\pm~$0.10702&$\pm~$0.10839&\\
  Te&-0.20893&-0.21228&-0.21136&-0.21306&&$\pm~$0.39572&$\pm~$0.38883&$\pm~$0.39790&$\pm~$0.39076&\\
  Bi(Sb)&0.40274&0.40161&0.40102&0.40156&&$\pm~$0.19922&$\pm~$0.19873&$\pm~$0.19846&$\pm~$0.19801&\\
  La(Y)&-0.39938&-0.39416&-0.40111&-0.39582&&$\pm~$0.29951&$\pm~$0.30219&$\pm~$0.29913&$\pm~$0.30120&\\
  $E_g$&0.12&0.12&0.15&0.28&&0.07&0.07& 0 &0.10  \\
\end{tabular}
\end{ruledtabular}
\end{table*}


{\it Method and crystal structure-} The electronic structures of LaBiTe$_3$,
YBiTe$_3$, LaSbTe$_3$ and YSbTe$_3$ are calculated in the framework of the
density functional theory with Perdew-burke-Ernzerbof type generalized gradient
approximation(GGA)\cite{perdew1996}. For our calculations, both the
\textit{Vienna ab initio simulation
package}(\textsc{vasp})\cite{kresse1993,kresse1996} with the projected
augmented wave method\cite{kresee1999} and the \textsc{bstate}
package\cite{fang2002} with plane-wave pseudo-potential method are employed. We
adopt the 12$\times$12$\times$12 k-point grid for self-consistent calculations.
The kinetic energy cutoff for the plane wave basis in BSTATE package is fixed
to 340 eV. SOC is included in all of our calculation, except in the
calculations for the ionic optimization which is confirmed to be
insensitive to SOC. 


LaBiTe$_3$, YBiTe$_3$, LaSbTe$_3$ and YSbTe$_3$ share the same rhombohedral
crystal structure with the space group D$^5$$_{3d}$(R$\overline{3}$m) (see
Ref.\onlinecite{madelung2000} and references therein). Here we take LaBiTe$_3$
as an example. Similar to Bi$_2$Te$_3$, the crystal consists of triangular
layers stacking along [111] direction with A-B-C$\cdots$ order. Five atomic
layers form a quintuple layer (QL) with the order of Te-X-Te-X$^\prime$-Te,
where X and X$^\prime$ denote La or Bi organized in certain pattern, as
discussed below. Though the lattice parameters have been measured by
experiments, up to our knowledge, the relative position between M and M' has
not been fully resolved experimentally. Thus we compare the total energy
obtained in {\it ab initio} calculation for different configurations to search
for the most stable structure. Our result indicates that the structures with La
and Bi separated in different atomic layers are
more stable than those with La and Bi mixed in the same atomic layer. 
There are two structures which have lowest energy. Because the energy
difference between the two configurations are very small, in the following we
will study both of them separately. (i) Type-I structure consists of stacking
quintuple layers along [111] direction, each of which is ordered as
Te-La-Te-Bi-Te, as shown in Fig. 1(a). This structure is inversion asymmetric.
(ii) Type-II structure consists of stacking pairs of quintuple layers along
[111] direction, ordered as Te-La-Te$^\prime$-La-Te and
Te-Bi-Te$^\prime$$^\prime$-Bi-Te. This structure is inversion symmetric. The
experimental lattice constants and optimized atomic coordinates for the two
structures are listed in Table I.

\begin{figure}
    \center
    \includegraphics[width=3.4 in]{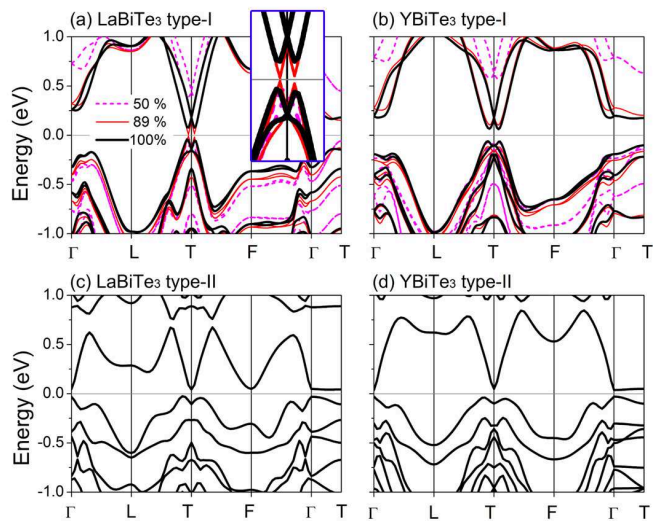}
    \caption{ Band structures for(a) LaBiTe$_3$ and (b) YBiTe$_3$ of type-I structures
    and (c) LaBiTe$_3$ and (d) YBiTe$_3$ of type-II structures. The strengths of spin-orbit
    coupling(SOC) are tuned to be 50\% (dashed purple lines), 89\% (thin red lines) and 100\%
    (thick black lines) respectively, in (a) and (b).
    The Fermi energy is set to zero. The energy gap decreases to zero and
    re-opens as SOC increases from zero to 100\% for LaBiTe$_3$, showing a clear
    band anti-crossing feature. The Dirac type level crossing at 89\% SOC is highlighted in
    the inset.
    (e) Brillouin zone for this class of materials. The four
    inequivalent time-reversal-invariant points are $\Gamma$(0;0;0), $L(\pi;0;0),
    F(\pi;\pi;0)$ and $T(\pi;\pi;\pi)$. The surface Brillioun zone is projected
    from the 3D bulk Brillouin zone into the plane of the atomic layer.}
    \label{fig:band1}
\end{figure}

\begin{figure}
   \begin{center}
      \includegraphics[width=3.5 in]{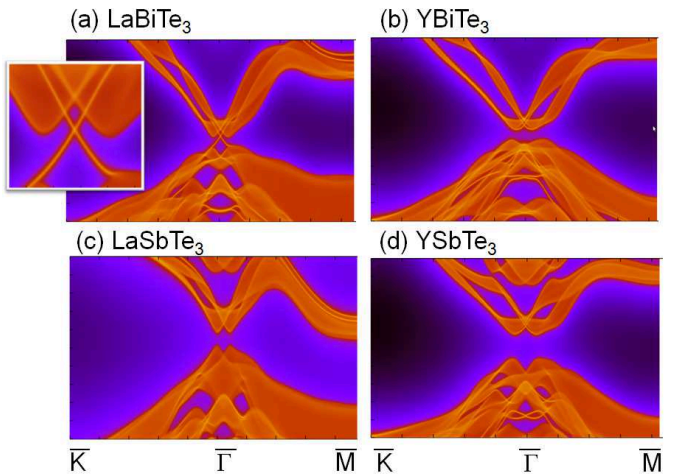}
    \end{center}
    \caption{ The band dispersion for an semi-infinite surface of
    (a) LaBiTe$_3$, (b) YBiTe$_3$, (c) LaSbTe$_3$ and YSbTe$_3$ of type-I structures. The inset
    of (a) highlights the surface states near the Dirac cone. }
    \label{fig:surfacestates}
\end{figure}


{\it Type-I structure.} We start by computing the band structures for
LaBiTe$_3$, YBiTe$_3$, LaSbTe$_3$ and YSbTe$_3$ in the type-I crystal
structure. As examples, Fig.2 shows the results for LaBiTe$_3$ and YBiTe$_3$.
Both of them have similar band gap 0.12 eV around the $T$ point. The band gap
of YBiTe$_3$ is close to the previous experimental value\cite{Rustamov1979}
0.18 $\sim$ 0.23 eV. As mentioned above, due to the lack of inversion symmetry,
the parity criteria for topological insulators\cite{fu2007a} does not apply.
Alternatively, we study the evolution of conduction and valence bands by tuning
the SOC strength artificially. Without SOC, any insulator must be topologically
trivial, so that a topological insulator phase can be identified by studying
the topological phase transition at some critical SOC strength\cite{bernevig2006d}. A Dirac type
level crossing between valence and conduction bands separate the topologically
trivial and nontrivial phases.\cite{bernevig2006d,zhang2009a} The continuous tuning of SOC is
obtained by changing the speed of light c in the \textsc{vasp} package. As
shown in Fig.2(a), upon the increase of SOC strength, the band gap of
LaBiTe$_3$ decreases and vanishes on a ring in Brillouin zone (BZ) near $T$ at
89\% SOC and opens again, which indicates a topological phase
transition\cite{murakami2007} and suggests that LaBiTe$_3$ is a topologically
non-trivial insulator with full SOC strength. In contrast, the band gap remains
finite upon turning on SOC for YBiTe$_3$, which is thus a topologically trivial
insulator. We perform the same calculations for LaSbTe$_3$ and YSbTe$_3$, and
find both of them are topologically trivial insulators.

The existence of topological surface states is one of the most important
physical consequence of the TIs. To further confirm the topological nature of
the proposed materials and study the detailed features of the surface states,
we obtain the dispersion of surface states based on {\it ab initio} method. We
construct Tight-binding(TB) Hamiltonian of the semi-infinite system with
Maximally localized Wannier functions(MLWF)\cite{marzari1997,souza2001} basis
for LaBiTe$_3$ family materials, and then apply the standard Green function
iteration method\cite{sancho1984,sancho1985} to obtain the surface density of
states. The detail of this method can be found in Ref.\onlinecite{zhang2009a}.
In Fig.3 we can see clearly that the topological surface states form a
single-Dirac-cone at the $\bar{\Gamma}$ point in 2D BZ within the bulk gap for
LaBiTe$_3$, while no such surface states are observed for YBiTe$_3$, LaSbTe$_3$
and YSbTe$_3$. This result determines the topologically non-trivial nature of
LaBiTe$_3$ without any doubt. From this calculation, we can further obtain the
Fermi velocity for LaBiTe$_3$ to be 4.0 $\times 10^5m/s$, similar to that of
Bi$_2$Se$_3$\cite{zhang2009}.

{\it Type-II structure.} The band structures for LaBiTe$_3$ and YBiTe$_3$ of
type-II are shown in Figs 2(c) and 2(d). Their band gaps are 0.07 eV near the
$\Gamma$ point, smaller than those of type-I structures. Due to inversion
symmetry, we can follow the parity criteria proposed by Fu and
Kane\cite{fu2007a} to calculate $\mathbb{Z}_2$ topological index. The product
of the parities of the Bloch wave functions are determined for the occupied
bands at all time-reversal-invariant momenta $\Gamma(0,0,0), L(\pi,0,0),
F(\pi,\pi,0), T(\pi,\pi,\pi)$ in BZ. For LaBiTe$_3$, we find that the product
of parities at $\Gamma$ is  `` + '' and those at $L, F, T$ are all  `` - ''.
Thus LaBiTe$_3$ is topological nontrivial with $\mathbb{Z}_2$ index (1;000).
For YBiTe$_3$ and YSbTe$_3$, we find that the products of parities at $\Gamma,
L, F, T$ are all `` - '', so that both of them are topologically trivial with
$\mathbb{Z}_2$ index (0;000). In addition, LaSbTe$_3$ is found to be a
semimetal. For both type-I and -II structures, it is noted that the states near
Fermi energy are partially composed by La/Y$-5 d_{z^2}$ as well as Bi/Sb-$6
p_z$ and Te-$5 p_z$ orbitals.


{\it Discussion-} In summary, our {\it ab initio} calculation has verified that
LaBiTe$_3$ is a strong topological insulator. The substitution of La in
LaBiTe$_3$ with other rare earth elements will open a wide platform to study
various topological non-trival behaviors. It is found that many isostructural
stoichiometric compounds similar to LaBiTe$_3$ also exist in the family of
ternary rare earth chalcogenides\cite{madelung2000}, such as LnBiTe$_3$ (with
Ln standing for La,Ce,Pr,......,Lu or Y) and LnSbTe$_3$, many of which have
intrinsic magnetic moments due to the localized $f$ electrons (e.g. Pr, Sm or
Gd). Since the strong correlation effect induced by $f$ electrons can not be
properly taken into account in the current GGA calculations, we will leave the
detailed study of these materials for the future work but discuss the possible
states induced by strong electron correlation. First, if the coupling between
magnetic moments is ferromagnetic, quantum anomalous Hall effect may emerge in
thin films or heterostructures of LnBiTe$_3$ when the band structure is normal
for one spin while inverted for the other, similar to the proposals in
Hg$_{1-x}$Mn$_x$Te\cite{liu2009c} and Bi$_{2-x}$Fe$_x$Se$_3$\cite{yu2010}.
Second, if the magnetic coupling is ferromagnetic within one atomic layer but
anti-ferromagnetic between two neighboring atomic layers, the dynamical axion
field may be realized, which is proposed as a novel optical modulators
device\cite{li2010}. The recently proposed topological Kondo
insulator\cite{dzero2010} may also exist in this class of materials due to the
existence of $f$ electrons. Besides the topological insulators the newly
proposed materials have another important potential application. As we have
shown, some materials in this family are trivial insulators such as YBiTe$_3$
and YSbTe$_3$. Due to their similarity with the Bi$_2$Te$_3$ family, the
trivial insulators in this new family are natural candidates of  barrier
materials for thin films or quantum wells of the TIs in the Bi$_2$Te$_3$ family
and the new LaBiTe$_3$ family. Especially, YBiTe$_3$ ($a$=4.46 \AA) -
LaBiTe$_3$ ($a$=4.39 \AA) and YBiTe$_3$ - Bi$_2$Te$_3$ ($a$=4.3835
\AA)\cite{madelung2000b} pairs have the best-matched lattice constants. Such a
barrier material may help protecting a clean surface of TI and greatly improve
the surface condition in transport experiments, which is a main difficulty in
the current experiments.

We would like to thank I. R. Fisher for the helpful discussion. This work is
supported by the Department of Energy, Office of Basic Energy Sciences,
Division of Materials Sciences and Engineering, under contract
DE-AC02-76SF00515. C.X.L. and B.Y. acknowledge financial support by the
Alexander von Humboldt Foundation of Germany. B.Y. thanks the support by the
Supercomputer Center of Northern Germany (HLRN Grant No. hbp00002).


\end{document}